\documentclass[12pt,a4paper]{article}
\usepackage{epsfig}
\begin{document}
\textwidth=135mm
 \textheight=200mm
\begin{center}
{\bfseries  Optimization of a large aperture dipole magnet for baryonic matter studies at Nuclotron}
\vskip 5mm
P.G.~Akishin$^{\dag}$, A.Yu.~Isupov$^{\dag}$, A.N.~Khrenov$^{\dag}$, P.K.~Kurilkin$^\dag$, V.P.~Ladygin$^{\dag,\ddag,}$\footnote{Corresponding author}, S.M.~Piyadin$^\dag$  and  N.D.~Topilin$^\dag$
\vskip 5mm
{\small {\it $^\dag$ Joint Institute for
Nuclear Research, 141980 Dubna, Russia}} \\
{\small {\it $^\ddag$ E-mail: vladygin@jinr.ru}}\\
\end{center}
\vskip 5mm
\centerline{\bf Abstract}
The aperture of the dipole magnet SP41 has been enlarged for the studies
of dense baryonic matter 
properties at Nuclotron. The homogeneity of the magnetic field in the
magnet centre has been improved. 
The measurement results of the magnetic field components and integral
are compared with results of 3D TOSCA calculations.
\vskip 10mm

\section{\label{sec:intro}Introduction}
The major goal in the studies of the dense baryonic matter at Nuclotron (BM@N project) \cite{bmn_CDR} is the measurement of strange and multi-strange baryons and mesons in 
heavy ion collisions  at the beam energies between 2 and 6 A$\cdot$GeV \cite{bmn_PoS}.  
The physics program can be extended to the measurements of the in-medium effects for 
strange particles decaying in hadronic modes \cite{brat1},
hard probes and correlations \cite{vasiliev_npps2011}, spin and polarization effects
\cite{lambda, bmn_dspin2013} etc. 

\begin{figure}[hbtp]
 \centering
  \resizebox{10cm}{!}{\includegraphics{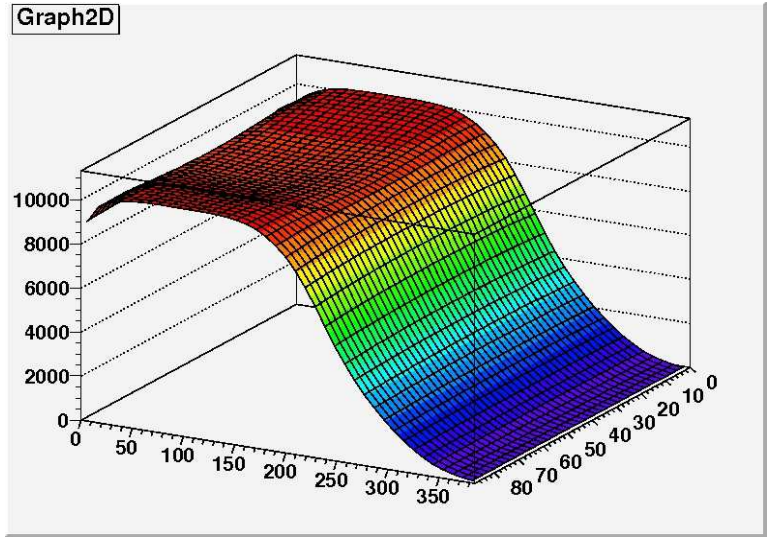}}
\caption{The vertical component $B_y$ of the magnetic field in the centre of
the  SP41 magnet before the modification. }
\label{fig:fig1}
\end{figure}

For these purposes an experimental set-up will be installed 
at the 6V beamline in the fixed-target hall of the Nuclotron.   
The 6V beamline contains  the quadrupole lenses doublet, two dipole magnets
allowing to
correct the beam position in the vertical and horizontal planes, and large aperture SP41 dipole magnet 
for the momentum measurements \cite{bmn_CDR}.  
The first results  with the relativistic deuteron \cite{terekhin_PoS} and
carbon \cite{piyadin_C12} beams are demonstrated the feasibility of the
dense baryonic matter studies with light nuclei using  6V  beamline infrastructure.

The modified SP41 dipole magnet will be used as an analyzing magnet \cite{bmn_CDR}. 
Initially, the length of the magnet pole along the beam was 2.50~m, width in the horizontal
direction was 1.70~m  and the height was  about 0.75~m.
This magnet had also a hole for the photo-camera in the 
upper pole since it was used previously for the experiments with streamer chamber. 
As the result the
magnetic field components are demonstrated the non-uniform behaviour. 
The $2D$ dependence of the magnetic field  vertical component $B_y$  of the non-modified magnet SP41
in $XZ$ plane is shown in Fig.\ref{fig:fig1}. 

However, the detection of multi-strange baryons requires the large aperture
silicon tracking system placement into the homogeneous magnetic field \cite{bmn_CDR,bmn_PoS,bmn_dspin2013}.
The SP41 magnet was modified to satisfy this requirement, namely,
the distance between the poles has been  
enlarged up to 1.05 m and the hole in the upper pole and horizontal
beams has been filled by the steel-15.
In this paper the results of the magnetic field measurements are presented for the 
modified dipole magnet  SP41. These results are compared with the 3D magnetic field TOSCA calculations. 
  
\section{\label{sec:TOSCA}Magnetic field calculations}

\begin{figure}[hbtp]
 \centering
  \resizebox{10cm}{!}{\includegraphics{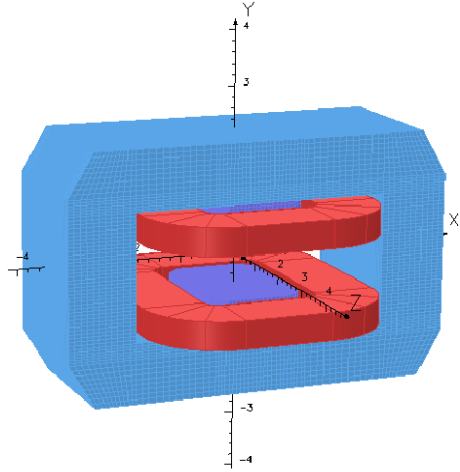}}
\caption{SP41 dipole magnet model for the magnetic field TOSCA \cite{tosca} 
calculations.}
\label{fig:fig2}
\end{figure}

\begin{figure}[hbtp]
 \centering
  \resizebox{8cm}{!}{\includegraphics{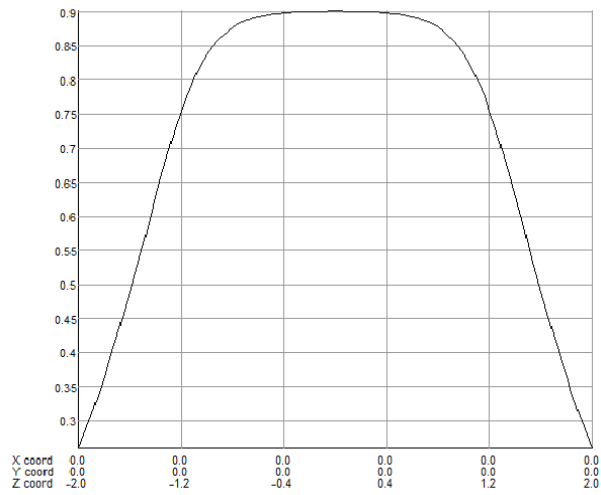}}
\caption{The 3D TOSCA calculation results for the magnetic field vertical component
$B_y$ in the centre of the modified magnet SP41 along the beam direction.}
\label{fig:fig3}
\end{figure}

The magnetic field calculations for modified SP41 dipole magnet 
has been performed using  3D  TOSCA code \cite{tosca}. 
Steel-10, steel-15 and copper were taken as the material for the yoke, magnet poles
and coils, respectively. The coordinate system was chosen as following:
$X$ axis is perpendicular to the beam direction in the horizontal plane, $Y$ is vertical and   $Z$ is along the beam and parallel to the magnet poles.  
The 3D model of the modified SP41 dipole magnet 
for the magnetic field TOSCA calculations is presented in
Fig.~\ref{fig:fig2}.

The 3D TOSCA calculation results for the magnetic field vertical component $B_y$
in the centre of the 
modified magnet SP41 along the beam direction are demonstrated in
Fig.~\ref{fig:fig3}. The current in the coils was taken as 1900~A. 
The maximal value of the vertical component $B_y$ was found  $\sim$0.9~T. The field integral is 
$\sim$2.9~T$\cdot$m, which is 
approximately 30~\% less than for the non-modified SP41 magnet.

\begin{figure}[htbt]
\begin{minipage}[t]{0.33\textwidth}
 \centering
  \resizebox{5cm}{!}{\includegraphics{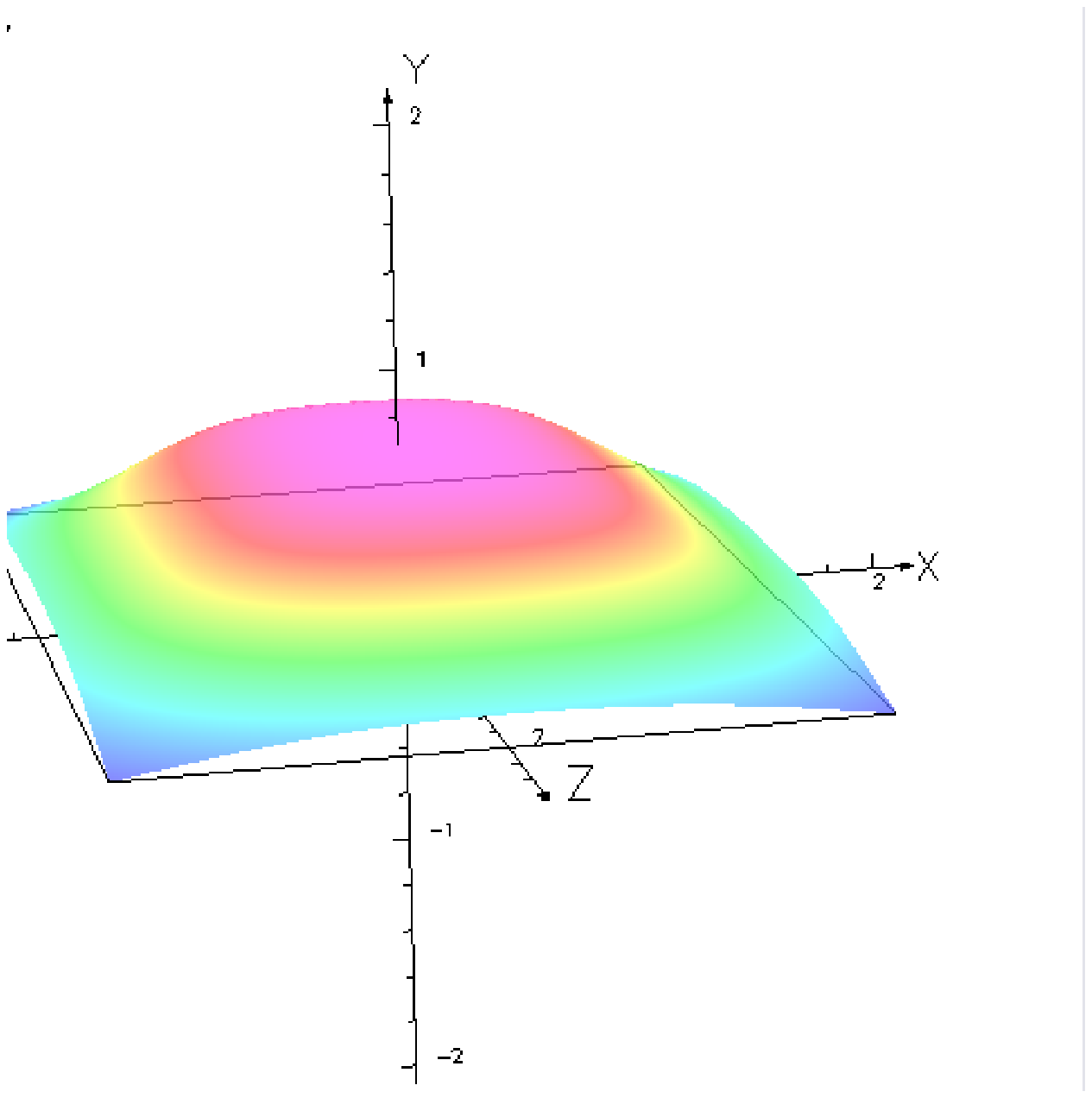}}
\end{minipage}\hfill\begin{minipage}[t]{0.33\textwidth}
 \centering
  \resizebox{5cm}{!}{\includegraphics{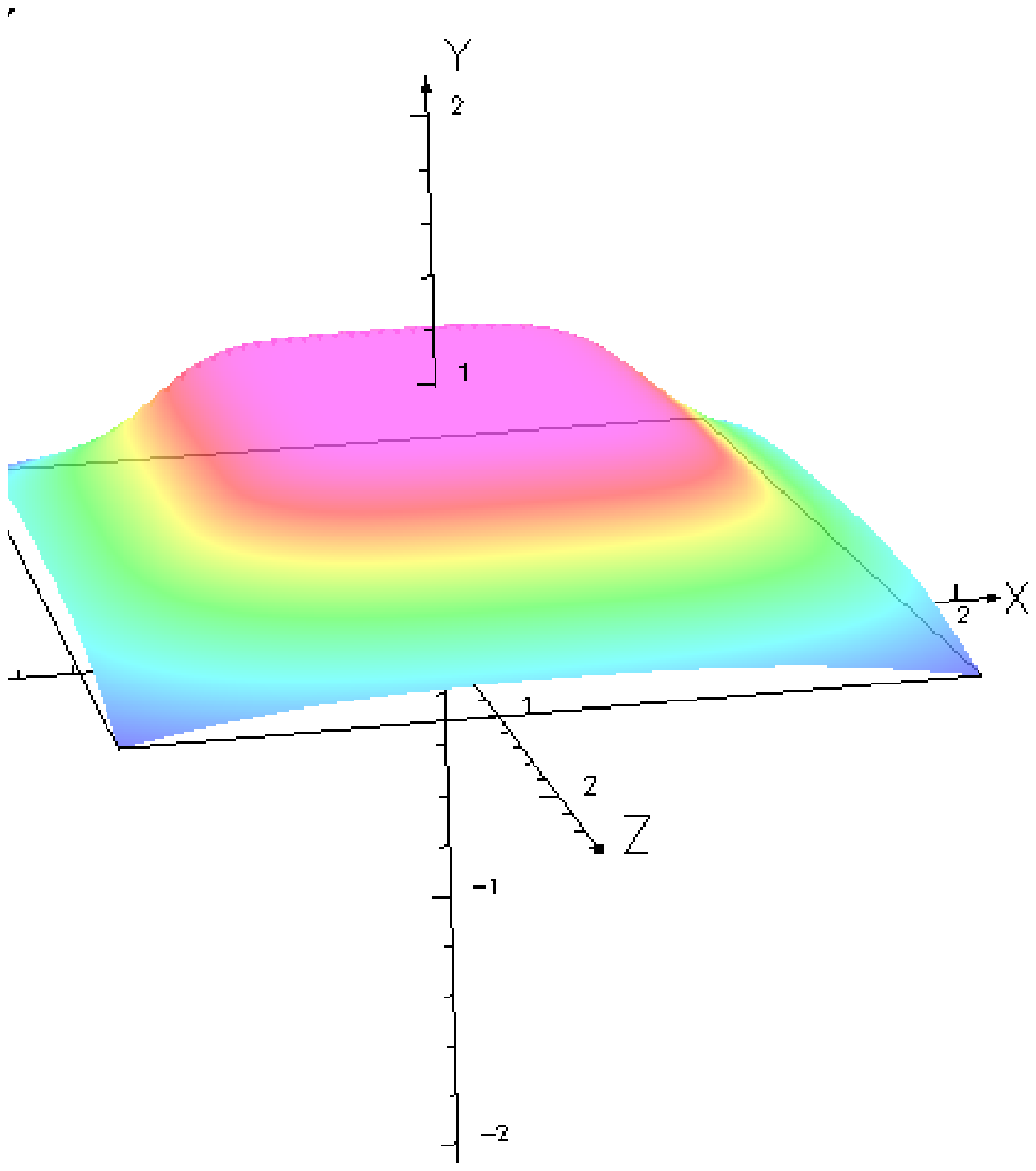}}
\end{minipage}\hfill\begin{minipage}[t]{0.33\textwidth}
 \centering
  \resizebox{5cm}{!}{\includegraphics{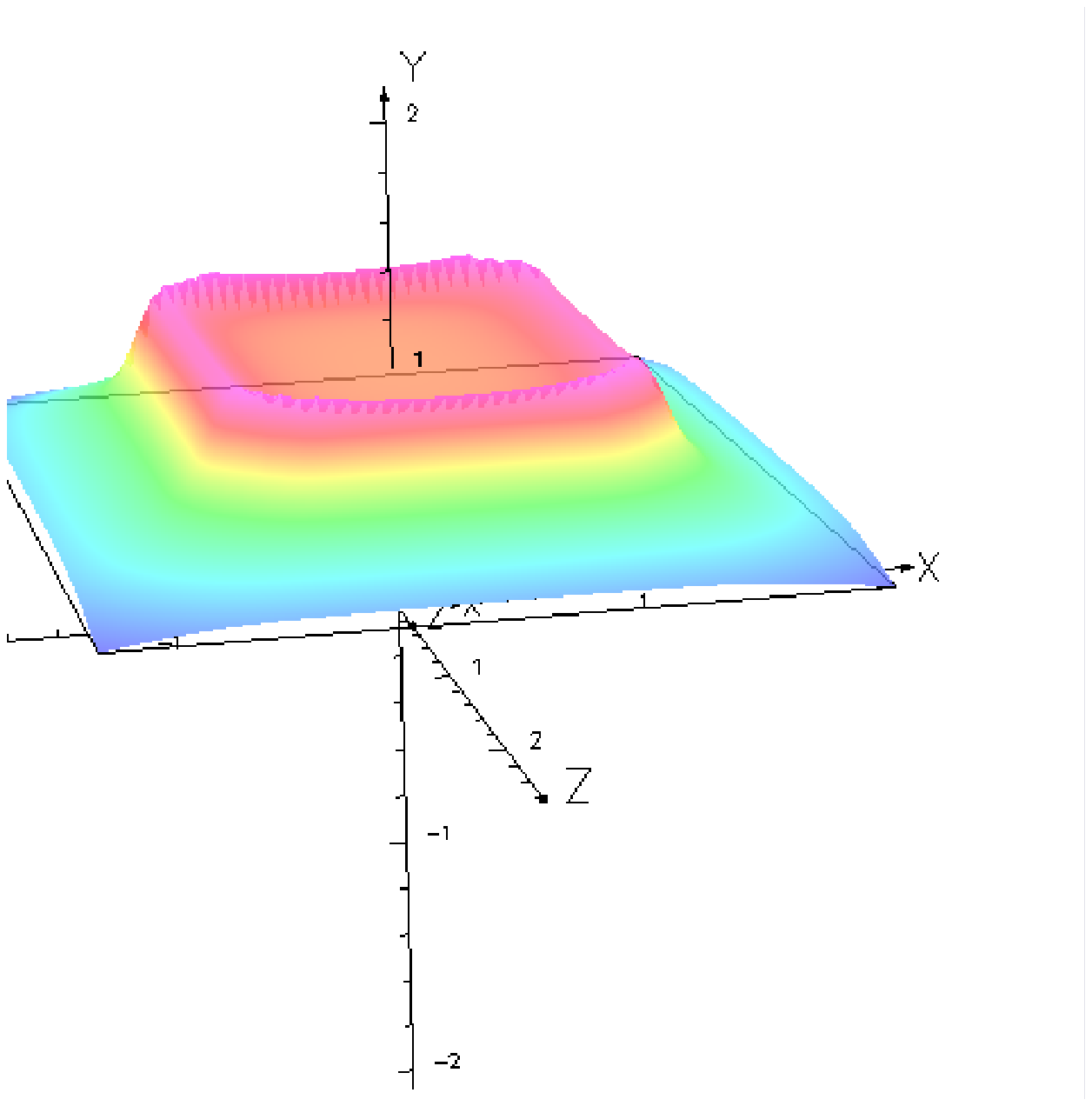}}
\end{minipage}
\caption{ 
From left to right: $2D$ distributions in $XZ$ plane of the magnetic field
vertical component $B_y$ of modified SP41 magnet at $Y = 0$~cm (centre of the
magnet), $Y = 30$~cm, and $Y = 50$~cm.
}
\label{fig:fig4}
\end{figure}

The $2D$ distributions in $XZ$ plane of the magnetic field vertical
component $B_y$   
of the modified SP41 magnet at $Y$=0 cm (centre of the magnet), $Y$=30 cm and
$Y$=50 cm are
presented in the left, middle, and right panels in Fig.~\ref{fig:fig4}.
The field demonstrates very smooth behaviour except the region around pole
of the magnet ($Y$=50 cm). The $3D$ results for $B_y$, $B_x$ and $B_z$
components of the magnetic field were incorporated
into the BM@N setup description for the simulation software \cite{bmn_dspin2013}.

\section{\label{sec:measur} Commissioning of the SP41 magnet}

\begin{figure}[hbtp]
 \centering
  \resizebox{10cm}{!}{\includegraphics{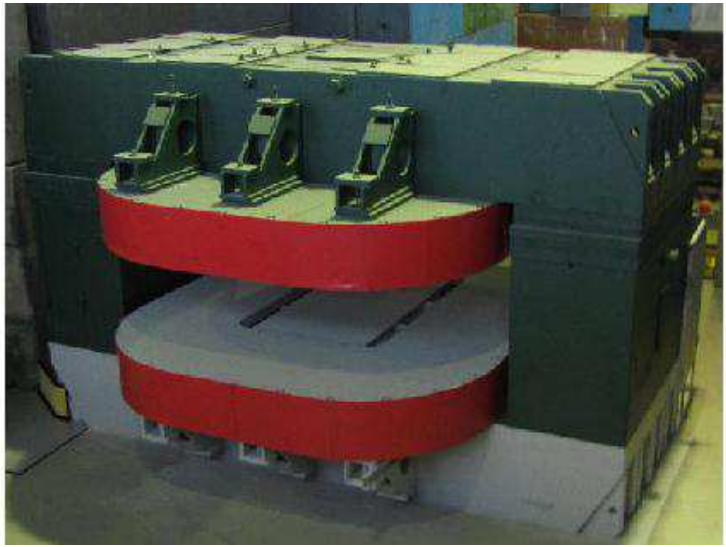}}
\caption{The view of modernized SP41 dipole magnet.}
\label{fig:fig5}
\end{figure}

During 2012--2013 the warm SP41 dipole magnet has been significantly modernized.
The magnet vertical gap has been enlarged by 30 cm up to 1.05 m. The upper pole and upper
horizontal beams have been filled by steel-15 in order to  improve the magnetic field
homogeneity. The upper coil (with renovated pole and horizontal beams) and 
lower coil alone have been rotated by 180$^\circ$
in the horizontal plane to provide optimal access to the magnet infrastructure  
and detectors inside the BM@N experimental zone \cite{bmn_CDR}. 
The magnet infrastructure, namely, pipes for cooling water,
the current leads and diagnostics have been also rotated by 180$^\circ$ and renovated.  
The view of modernized SP41 dipole magnet is presented in Fig.\ref{fig:fig5}.

The commissioning of the SP41 magnet has been performed in two steps.
Firstly, the operation of the magnet with the 1650 A current 
in the coils  has been checked several times during several hours.
Second step includes the measurements of the magnetic field  
in the centre of the magnet, the magnetic field integral and  magnetic fringe field within 
working range of the current in the coils (up to 1960 A).

\begin{figure}[hbtp]
 \centering
  \resizebox{10cm}{!}{\includegraphics{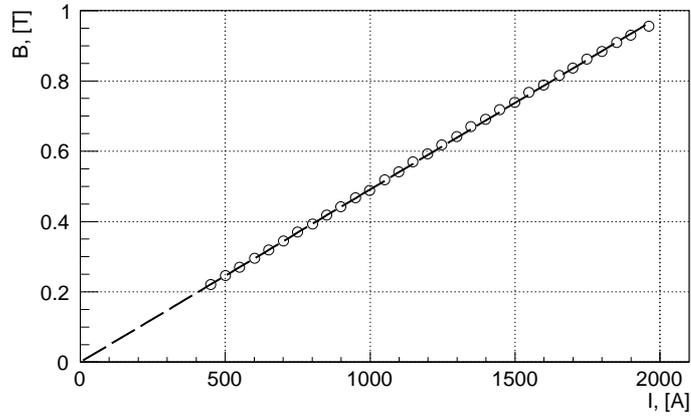}}
\caption{The
$B_y$ vertical component of the magnetic field 
in the centre of modernized SP41 dipole magnet
as a function of the  current in the coils \cite{rukoyatkin}. The dashed curve is the result of the quadratic function approximation.}
\label{fig:fig6}
\end{figure}

\begin{figure}[hbtp]
 \centering
  \resizebox{10cm}{!}{\includegraphics{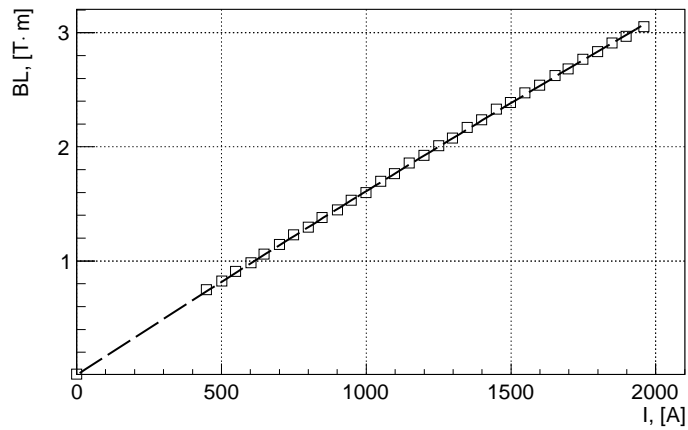}}
\caption{The
magnetic field integral $\int Bdl$  of modernized SP41 dipole magnet
as a function of the  current in the coils \cite{rukoyatkin}. The dashed curve is the result of the quadratic function approximation.}
\label{fig:fig7}
\end{figure}

The measurements of the $B_y$ vertical component of the magnetic field  
in the centre of modernized SP41 dipole magnet
($X=0$, $Y=0$, $Z=0$) have been performed using planar Hall probe \cite{rukoyatkin}. 
The
$B_y$ in the centre of modernized SP41 dipole magnet
as a function of the current in the coils is shown in Fig.\ref{fig:fig6}. 
The dashed curve is the result of the  quadratic function approximation. 
The $B_y$ demonstrates the linear dependence on the current with negligible contribution
of the quadratic term.  The  $B_y$ at 1900 A  is equal to 0.933$\pm$0.003 T
being in good 
agreement with the 3D TOSCA calculation shown in Fig.\ref{fig:fig2}. 

The measurements of the magnetic field integral $\int Bdl$  of modernized SP41 dipole magnet has been 
performed using the method of the current-carrying filament \cite{rukoyatkin}. 
The
magnetic field integral $\int Bdl$  of modernized SP41 dipole magnet
as a function of the  current in the coils is shown in Fig.\ref{fig:fig7}. 
The dashed curve is the result of the quadratic function approximation.  
The magnetic field integral $\int Bdl$ is equal to 2.986$\pm$0.009 T$\cdot$m at 1900~A. 
The non-linear  contribution is about 5\%. The measured value is in good accordance
with the  3D TOSCA calculation results \cite{bmn_CDR}. 
The field integral provides the bending angle of $\sim$70 mrad for 
charged particles with $p/Z$=13 GeV/$c$. 
On the other hand, since the saturation effect for the magnetic 
field integral is not observed  at 1900~A, the current value in the coils can be increased. This
will provide larger bending angle and better momentum resolution.     
 
\begin{figure}[hbtp]
 \centering
  \resizebox{10cm}{!}{\includegraphics{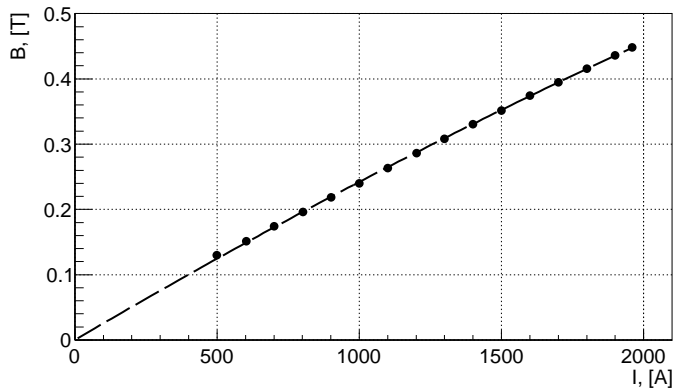}}
\caption{The
magnetic field $|B|$ measured with the planar 3D Hall probe  \cite{bmn_CDR}
at $X=0.2$~m, $Y=-0.22$~m, $Z=1.6$~m
as a function of the  current in the coils. 
The dashed curve is the result of the quadratic function approximation.}
\label{fig:fig8}
\end{figure}

\begin{figure}[hbtp]
 \centering
  \resizebox{10cm}{!}{\includegraphics{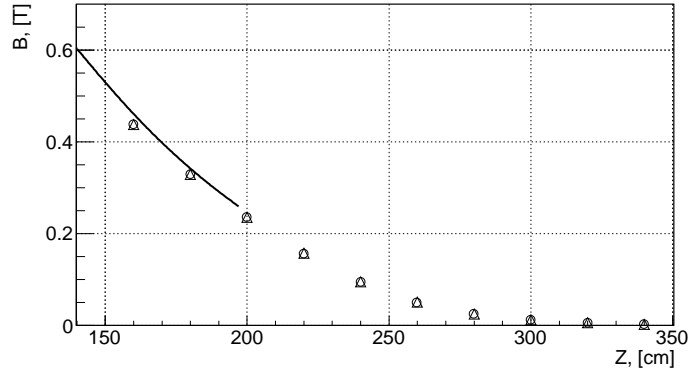}}
\caption{The
magnetic field $|B|$ for the 
current value in the coils of 1900~A as a function of $Z$ coordinate  ($X=0.2$~m and  $Y=-0.22$~m).
The open circles and triangles are the data obtained with the planar and coaxial 3D Hall probes \cite{bmn_CDR},
respectively. 
The solid curve is the result of 3D TOSCA calculation scaled by factor 0.92.}
\label{fig:fig9}
\end{figure}

The measurements of the magnetic field  $|B|$ at $X=0.2$~m, $Y=-0.22$~m, $Z=1.6$~m   
have been performed using planar 3D Hall probe described in details in ref.\cite{bmn_CDR}.  
The position of this point 
is out of the magnet poles. 
The
measured magnetic field $|B|$  
as a function of the current value in the coils is shown in Fig.\ref{fig:fig8}. 
The dashed curve is the result of the  quadratic function approximation. 
The $|B|$ demonstrates the non-negligible contribution
of the quadratic term at large values of the current in the coils.  
For instance, the non-linear  contribution is about 12\% at the current value of 1900~A.
Therefore, the fringe field demonstrates the saturation effect at large 
currents in the coils, which is absent for the field in central region of the magnet.

The measurements of the fringe magnetic field  as a function 
of the distance from the magnet centre $Z$ at fixed values of
$X=0.2$~m and  $Y=-0.22$~m coordinates have been performed using planar and coaxial 3D Hall probes \cite{bmn_CDR}.
The
magnetic field $|B|$ for the 
current value in the coils of 1900~A as a function of $Z$ coordinate  is shown in Fig.\ref{fig:fig9}.
The open circles and triangles represent the data obtained with the planar and coaxial 3D Hall probes \cite{bmn_CDR},
respectively.  One can see that
the results obtained by these two probes are in good agreement.  
The solid curve is the result of 3D TOSCA calculation scaled by factor 0.92.  
The shape of the measured magnetic field $|B|$ reproduces the TOSCA results.
The observed difference can be recovered by the optimization of  
the SP41 dipole magnet model for the magnetic field TOSCA \cite{tosca} 
calculations.

\section{\label{sec:conclusions} Conclusions}

\begin{itemize}

\item 
The SP41 dipole magnet has been  modernized for the studies of dense baryonic matter 
properties at Nuclotron. Namely,  
the magnet vertical gap has been enlarged  up to 1.05 m to increase the angular acceptance for
detection of hyperons \cite{bmn_PoS, lambda, bmn_dspin2013},  
the magnetic field homogeneity improvement by the filling of the existed hole in the upper 
pole and horizontal beams  by steel-15  has been achieved,
the renovation of the magnet infrastructure has been made.
\item
The magnetic field in the centre of the magnet and out of the poles as well as field integral 
have been measured  as a function of current value in the coils. The 
value of the magnetic field in the magnet centre and field integral at 1900~A
are in good agreement with the results of 3D TOSCA calculation,
while the measured fringe field at the same current is several percents below
the calculated one.
\item 
Further steps are the measurements of the magnetic field components along the
optical axis of the SP41 magnet and 3D mapping of the magnetic field. 
\end{itemize}
\vspace{0.5cm}
The authors thanks the technical services of LHEP participated in the modification
of the SP41 dipole magnet, especially, G.S.~Berezin, I.Ya.~Nefediev, A.V.~Shabunov, V.I.~Sharapov.   
The authors are grateful to P.A.~Rukoyatkin for the magnetic measurements,  I.A.~Bolshakova and S.~Timoshin for their assistance in
the use of 3D Hall probes.  The work has been supported  in part by RFBR 
under grant $N^o$13-02-00101a.

\end{document}